\documentclass[journal = jctcce]{achemso}  
\usepackage{amssymb}
\usepackage{amsmath}
\usepackage{graphicx}
\usepackage{multirow}
\usepackage{comment}
\usepackage{longtable}
\usepackage{color}
\usepackage{xfrac}
\usepackage{threeparttable} 
\graphicspath{{./fig/}}

\newcommand{\eqn}[1]{\mbox{Eq.\hspace{1pt}(\ref{#1})}}

\newcommand{\eqtn}[2]{\begin{equation} \label{#1} #2 \end{equation}}

\def\br{{\mathbf{r}}}

\def\etal{{\it et al.}}

\def\vw{{von Weizs\"{a}cker}}

\title{GGA-Level Subsystem DFT Achieves Sub-kcal/mol Accuracy Intermolecular Interactions by Mimicking Nonlocal Functionals}
\author{Xuecheng Shao}
\email{xuecheng.shao@rutgers.edu}
\affiliation{Department of Chemistry, Rutgers University, Newark, NJ 07102, USA}
\author{Wenhui Mi}
\email{wenhui.mi@rutgers.edu}
\affiliation{Department of Chemistry, Rutgers University, Newark, NJ 07102, USA}
\author{Michele Pavanello}
\email{m.pavanello@rutgers.edu}
\affiliation{Department of Chemistry, Rutgers University, Newark, NJ 07102, USA}
\affiliation{Department of Physics, Rutgers University, Newark, NJ 07102, USA}

\begin{document}
\maketitle
\begin{abstract}
	The key feature of nonlocal kinetic energy functionals is their ability to reduce to the Thomas-Fermi functional in the regions of high density and to the \vw\ functional in the region of low density/high density gradient. This behavior is crucial when these functionals are employed in subsystem DFT simulations to approximate the nonadditive kinetic energy. We propose a GGA nonadditive kinetic energy functional which mimics the good behavior of nonlocal functionals retaining the computational complexity of typical semilocal functionals. The new functional reproduces Kohn-Sham DFT and benchmark CCSD(T) interaction energies of weakly interacting dimers in the S22-5 and S66 test sets with a mean absolute deviation well below 1 kcal/mol.
\end{abstract}
\newpage
\section{Introduction}
Subsystem DFT (sDFT) is a divide-and-conquer method that extends the applicability of conventional DFT to system sizes larger than typically accessible \cite{neug2009b,weso2015,jaco2014,krish2015a}. The main idea behind sDFT is to split the noninteracting kinetic energy into subsystem additive and nonadditive parts \cite{grit2013,weso2006,sen1986,cort1992,weso1993,gord1972}. Namely,
	\eqtn{nad}{T_s[\rho] = \sum_I^{N_S}T_s[\rho_I] + T_s^{nadd}[\{\rho_I\}],}
where $\{\rho_I(\br)\}$ is the set of subsystem electron densities. In doing so, orthogonalization of wavefunctions belonging to different subsystems is completely bypassed \cite{good2010,fderelease,jaco2008b,Unsleber_2018,H_fener_2020}, and a formally linear-scaling behavior can be attained (or quasi-linear due to the need to solve for Coulomb interactions which scales like $\mathcal{O}(N\log(N)$). Naturally, for the method to be accurate, the additive terms are treated exactly (i.e., the kinetic energy internal to the subsystems is evaluated with the subsystem orbitals) while to improve efficiency, the nonadditive functional is defined simply through \eqn{nad} and is evaluated with pure density functionals. Clearly, the accuracy of the method is directly related to the quality of the pure functionals employed as nonadditive functionals \cite{weso2003,weso1997b,Goez_2017,fux2010,gotz2009}. 

The common strategy in sDFT is to borrow already available pure density functionals for the noninteracting kinetic energy and use them in \eqn{nad} to obtain approximations for the nonadditive kinetic energy functional (NAKE). LDA, GGA and nonlocal NAKEs have been proposed by this strategy. GGA NAKEs already provide a semiquantitative description of inter-subsystem interactions (molecule-molecule and molecule surface) \cite{gotz2009,genova2014,schl2015}. However, only recently nonlocal functionals (i.e., functionals that depend on the density evaluated  simultaneously in two or more separate points in space) have been employed as NAKE with great success \cite{Mi_2019,kevo2014b,sinh2015}. Such GGA nonadditive functionals as revAPBEK and LC94 are expected to perform with an accuracy averaging slightly above 1 kcal/mol, however, the shape of the energy curves are incorrectly too attractive (see refs.\ \citenum{Mi_2019} and \citenum{schl2015}, and {\it vide infra}). When these functionals are corrected by a fitted repulsive function (in the same spirit as the repulsive part of a Grimme's D$n$ correction) the shape of the energy curves involved are significantly improved \cite{schl2015}. Nonlocal functionals also make very good NAKEs, and further improve the performance of GGA NAKEs in sDFT simulations to a MAD below 1 kcal/mol for the S22-5 set, with particularly strong improvement for strained structures, and correctly reproducing the shape of the energy curves\cite{Mi_2019}.


The question we address in this work is: can we achieve an accuracy similar to the one of nonlocal functionals with a GGA functional? Since the early work on GGA NAKEs \cite{tran2002b}, it became clear that in order to satisfy the non-negativity condition, $T_s^{nadd}\geq 0$, it is important to use the Thomas-Fermi functional, $T_{\rm TF}[\rho]=\int \rho^{5/3}(\br)d\br$, for low values of the reduced gradient, $s(\br)=|\nabla\rho(\br)|/(2\rho(\br)k_F(\br))$, where $k_F=(3\pi^2\rho(\br))^{1/3}$ is the local Fermi wavevector. To easily achieve this, an enhancement factor, $F_s$, is introduced. Namely,
\eqtn{GGA}{T_s^{GGA}[\rho] = \int \tau_{\rm TF}(\br) F_s(s(\br))d\br,}
where $\tau_{\rm TF}(\br)$ is the Thomas-Fermi (TF) kinetic energy density. Such a strategy is commonplace. It is found in, for example, the revAPBEK \cite{lari2011} kinetic energy functional as well as many others \cite{tran2002b,revAPBEK,Karasiev_2006}. They share the following approximate form of the enhancement factor borrowed from the PBE exchange functional \cite{PBEc},
\eqtn{pbek}{F_s^{approx}(s)=1+\frac{\kappa s^2}{1+\frac{\kappa}{\mu}s^2}.}
Laricchia \etal\ \cite{lari2011} found that revAPBEK outperformed all other GGA functionals. Thus, in this work we use revAPBEK as a reference for the performance of GGA functionals.

Nonlocal functionals achieve the goal of functional positivity by providing the following ansatz for the kinetic energy functional\cite{huan2010,wang2000},
\eqtn{nl}{T_s[\rho]=T_{\rm TF}[\rho]+T_{\rm vW}[\rho]+T_{NL}[\rho],}
where $T_{\rm vW}$ is the \vw\ kinetic energy (which violates positivity \cite{Jiang_2018}) and $T_{NL}$ is the nonlocal contribution. $T_{NL}$ is designed to cancel out $T_{\rm TF}$ in the tail of atoms where $T_{\rm vW}$ is expected to be accurate, and to cancel out $T_{\rm vW}$ in the regions of high density/low density gradient where $T_{\rm TF}$ is expected to be accurate. Thus, it is not a surprise that nonlocal nonadditive functionals are the currently best-performing NAKEs \cite{Mi_2019}.

\section{The SMP functional}

Cognizant of the analysis above, we set out to develop a NAKE that takes inspiration from the success of nonlocal functionals while targeting the computational complexity of GGA functionals. We term the new functional SMP. SMP's ``parent'' kinetic energy functional features an energy density that is a mix of two functionals,
\begin{equation}
\begin{aligned}
	T_{s}[\rho]&=\int \tau[\rho](\br) d\br \\
	&=\int W[\rho](\br) \tau_{\rm TF} [\rho](\br) + \bigg(1-W[\rho](\br)\bigg)\tau_{GGA}[\rho](\br) d\br,
\end{aligned}
\end{equation}
where $\tau[\rho]$, $\tau_{GGA}[\rho]$ are the total kinetic energy density and the one of a GGA functional, $T_{GGA}[\rho]$ (we use revAPBEK in this work), respectively. Mixing functionals is not a new concept, especially for the xc functional \cite{Ghassemi_2019,Tchakoua_2019,Smeets_2019,Peng_2016,Mardirossian_2015}. $ W[\rho]$ is a switching function depending on a constant parameter,  $\rho^*$, as follows,
\begin{equation}
	\label{w}
	W[\rho]= \begin{cases}
		\rho(\br)/\rho^* &\text{if } \rho < \rho^*, \\
		1 &\text{if } \rho \geq \rho^*.
	\end{cases}
\end{equation}

The corresponding kinetic potential is
\begin{equation}
\begin{aligned}
	v_{s}[\rho](\br)& =\frac{\delta T_{s}[\rho]}{\delta \rho(\br)} \\
	& = W[\rho](\br) v_{\rm TF}[\rho](\br) + \frac{\partial W[\rho]}{\partial \rho(\br)} \tau_{\rm TF}[\rho](\br) \\
	& +\left[1-W[\rho](\br)\right] v_{GGA}[\rho](\br)- \frac{\partial W[\rho]}{\partial \rho(\br)} \tau_{GGA}[\rho](\br),
\end{aligned}
\end{equation}
where $v_{\rm TF}[\rho](\br) =\frac{\partial \tau_{\rm TF}[\rho]}{\partial \rho(\br)}$, $v_{GGA}[\rho](\br) =\frac{\partial \tau_{GGA}[\rho]}{\partial \rho(\br)} -\nabla \cdot \frac{\partial \tau_{GGA}[\rho]}{\partial \nabla \rho(\br)}$, and

\begin{equation}
	\frac{\partial W[\rho]}{\partial \rho(\br)} = \begin{cases}
		1/\rho^* &\text{if } \rho < \rho^*,  \\
		0 &\text{if } \rho \geq \rho^*.
	\end{cases}
\end{equation}
The final kinetic potential can be expressed as,
\begin{equation}
	v_{s}[\rho](\br) = 
	W[\rho](\br)v_{\rm TF}[\rho](\br) + \frac{\tau_{\rm TF}[\rho](\br)}{\rho^*} + \begin{cases}
	\left[1-W[\rho](\br)\right]v_{GGA}[\rho](\br) -\frac{\tau_{GGA}[\rho](\br)}{\rho^*}  &\text{if } \rho < \rho^*, \\
	0 &\text{if } \rho \geq \rho^*.
\end{cases}
\end{equation}


So far, the above equations are appropriate for an approximation for the noninteracting kinetic energy. To obtain a NAKE, we can use the decomposable NAKE formula \eqn{nad}. For SMP the $\rho^*$ parameter is crucial and it is chosen following the steps below:
\begin{enumerate}
	\item Compute the ratio between $\rho_I(\br)$ and $\rho(\br)-\rho_I(\br)$. Namely,
		\eqtn{ovlp}{S_{I}(\br)=\frac{\rho_I(\br) }{\rho(\br)-\rho_I(\br)}.}
	\item Special points $\br_h$ are those where $S_I(\br_h)=1$.
	\item $\rho^*$ is given by the following expression,
		\eqtn{rhostar}{\rho^* = \max_{\br_h}\left\{ \rho_I(\br_h) + \rho(\br_h) \right\}.}
\end{enumerate}
The parameter $\rho^*$ needs no adjustment during a single-point energy calculation and is unique to each subsystem. 
The total kinetic energy functional is evaluated with a $\rho^*$ obtained by averaging the $\rho^*$ of all the subsystems.

$\rho^*$ is determined at the beginning of the calculation and kept constant throughout the Self-Consistent Field procedure. This can be potentially problematic during an ab-initio molecular dynamics (MD). However, we plan to determine $\rho^*$ from the electron densities of the last MD step avoiding the need to run two single point calculations (i.e., one for $\rho^*$ and one with SMP employing the determined $\rho^*$). In this work, for simplicity we determine $\rho^*$ by performing a preliminary calculation with the revAPBEK NAKE. To dispel doubts about the robustness of the method, we also verified that determining $\rho^*$ self-consistently (i.e., re-running calculations updating $\rho^*$ until convergence) leads to essentially the same results (see supplementary document Figure S1).

Let us analyze \eqn{rhostar} to make sense of the proposed procedure.  Firstly, the special points $\br_h$ select regions of space where there is large overlap between the electron densities of the subsystems. A NAKE should have an awareness of the inter-subsystem density overlap \cite{last2008,jaco2013,jaco2014,Jiang_2018}. Thus, $\rho^*$ depends on the value of the subsystem density at the special points to encode such a dependence. Secondly, the choice of $\rho^*$ should be such that the switching function in \eqn{w} correctly weighs the regions of high density (\emph{relative} to the inter-subsystem overlap) assigning the TF functional to these regions and a GGA functional to the regions of \emph{relative} low  density. 

\section{Computational Details}

The molecules are placed in an orthorhombic box, which is large enough so that the nearest-neighbor periodic images are more than 12 \AA apart. This ensures that the interactions between the studied systems and their periodic images are negligible. All sDFT calculations in this work were performed with a development version of the in-house eDFTpy Python-based density embedding software \cite{shao2021edftpy}. All KS-DFT benchmark calculations were performed with the Quantum ESPRESSO (QE) package \cite{qe_new}. In both subsystem DFT and KS-DFT calculations, the Perdew--Burke--Ernzerhof (PBE) form of the GGA xc functional is employed \cite{PBEc}. The GBRV ultrasoft pseudopotentials \cite{garr2014} are adopted due to their excellent transferability. The kinetic energy cutoffs of wave-function and density are 70 and 400 Ry, respectively. The convergence threshold for self-consistent calculations is $ 1\times10^{-8} $ Ry.

\section{Results and Discussion}

In Table \ref{tab:energy} we list mean absolute deviations (MADs) of the interaction energies computed with sDFT and various NAKEs against the KS-DFT result employing the PBE xc functional. The point of this comparison is to inspect how well the NAKEs perform against KS-DFT which employs the exact noninteracting kinetic energy. This is a meaningful comparison \cite{gotz2009,schl2015} because if the NAKE is exact, the KS-DFT result should be recovered.

\begin{table}[htp]
	\caption{\label{tab:energy}Summary of the Mean absolute deviations (MADs) of the interaction energies computed with sDFT carried out with TF, revAPBEK, LMGPA and SMP NAKEs against KS-DFT results. All values are given in kcal/mol.}
\begin{tabular}{cccccc}
	\hline
	\hline
	         Set           & Functional & Hydrogen & Dispersion & Mixed & Total \\ \hline
	\multirow{4}{*}{S22-5} &     TF     &   4.40   &    1.60    & 1.32  & 2.40  \\
	                       &  revAPBEK  &   0.82   &    2.01    & 0.53  & 1.16  \\
	                       &   LMGPA    &   0.97   &    0.89    & 0.60  & 0.82  \\
	                       &    SMP     &   0.81   &    0.81    & 0.17  & 0.60  \\ \hline
	 \multirow{4}{*}{S66}  &     TF     &   5.08   &    2.48    & 2.14  & 3.28  \\
	                       &  revAPBEK  &   0.58   &    1.67    & 0.91  & 1.06  \\
	                       &   LMGPA    &   1.28   &    1.24    & 0.49  & 1.03  \\
	                       &    SMP     &   1.04   &    0.33    & 0.13  & 0.52 \\
	                       \hline
	                       \hline
\end{tabular}
\end{table}

As we can see from the table, SMP delivers almost always the most accurate interaction energies for the two data sets chosen (S66 and S22-5) with exception for hydrogen bonded systems. It is generally known that GGA NAKEs like revAPBEK and LC94 are particularly good with hydrogen bonded systems \cite{gotz2009,kevo2006} leading these functionals to describe liquid water very accurately \cite{Genova_2016a}. The reason for the good behavior of GGA NAKEs for hydrogen bonded systems is because the error introduced in the NAKE offsets errors brought in by the nonadditive part of the exchange-correlation (xc) functional\cite{weso2003,weso2002,gotz2009,tran2001} (e.g., the long range part of the xc functional is known to include errors of self-interaction \cite{Ruzsinszky_2006}). SMP, however, is overall of higher quality compared to these GGA functionals, and performs even better than a nonlocal functional with density-dependent kernel (LMGPA) \cite{mi2018nonlocal,mi2019LMGP}.

\begin{figure}[htp]
	\begin{tabular}{cc}
	\includegraphics[width=0.5\linewidth]{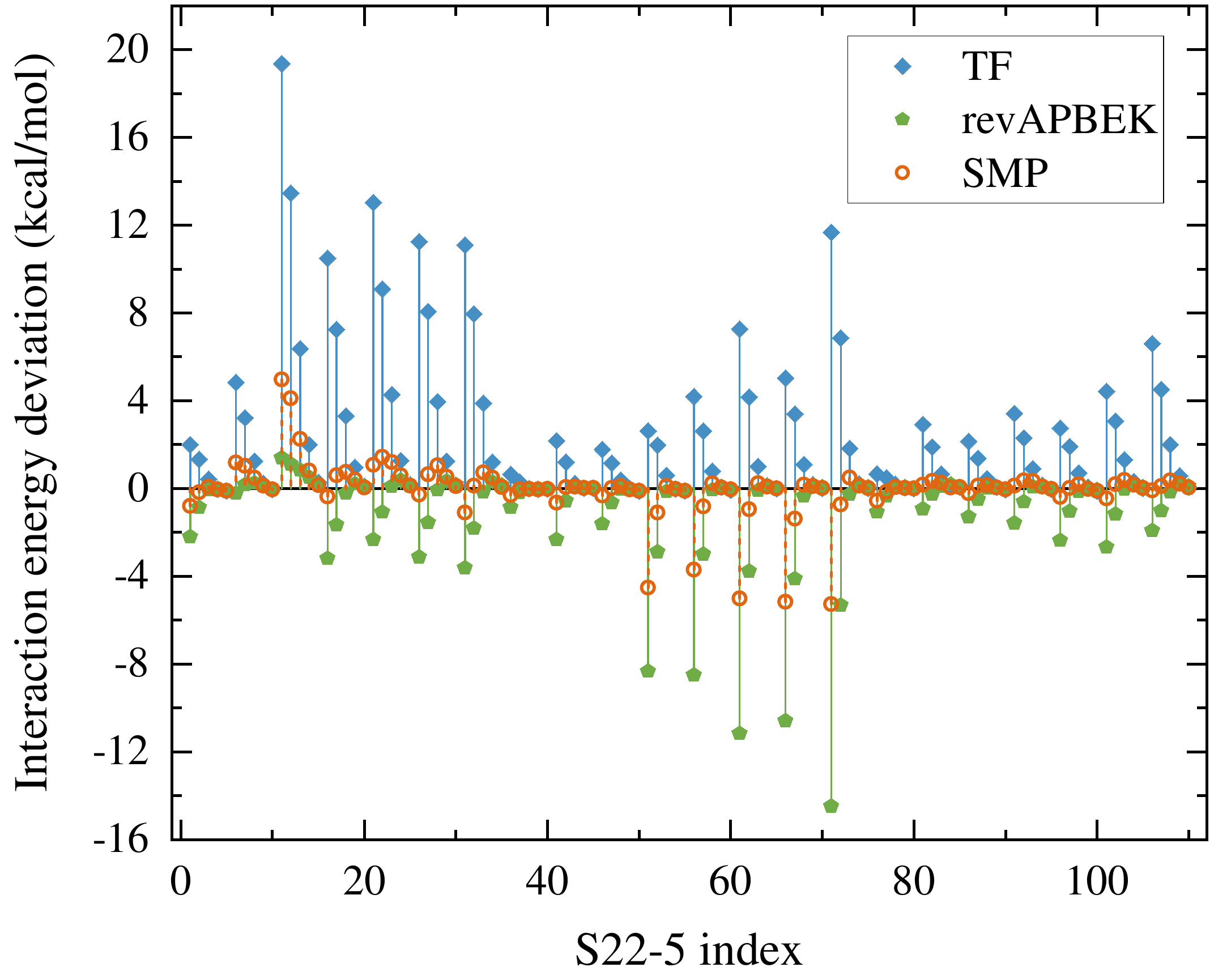} &
	\includegraphics[width=0.5\linewidth]{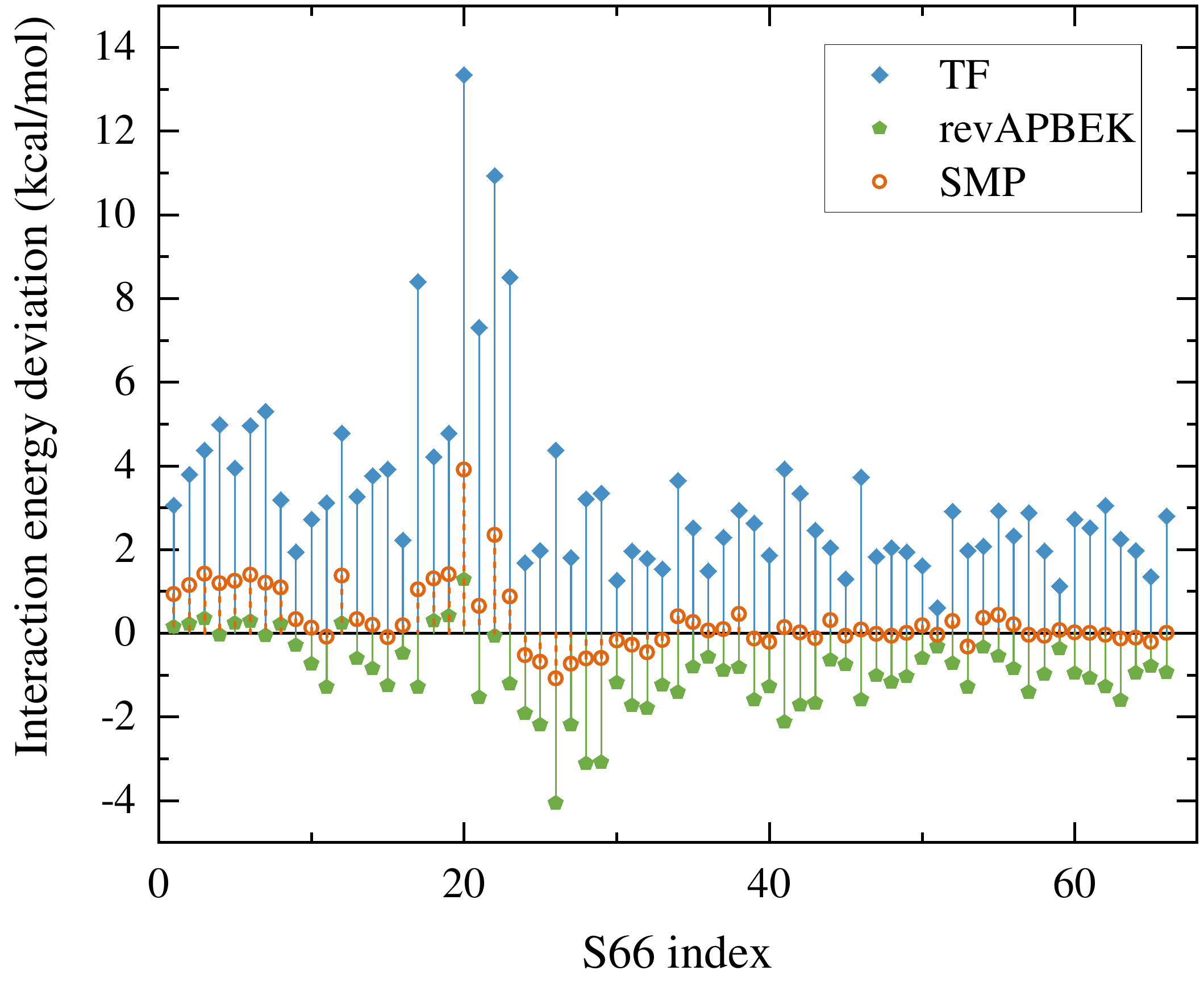}
	\end{tabular}
	\caption{\label{fig:energy} Interaction energy deviations in kcal/mol between sDFT with TF, revAPBEK and SMP KEDFs against the KS-DFT results for the S22-5 and S66 test sets. Both KS-DFT and sDFT calculations are carried out with the PBE xc functional.}
\end{figure}

In Figure \ref{fig:energy}, we break down each contributing dimer to the S22-5 and S66 test sets. From the figure, it is evident why SMP works as NAKE. SMP effectively interpolates between TF and a GGA functional. GGA functionals, such as revAPBEK, tend to underestimate the binding energy while TF always overestimates the binding energy. An interpolation of the two will naturally improve the binding energy values. However, SMP not only exploits error cancellation to deliver improved binding energies but also does a better job in reproducing the electron densities as we explain below.

The next test is to check the accuracy in the predicted electron density. Once again we carry out this test against KS-DFT densities. In Table \ref{tab:density}, we showcase the density deviations, computed as $\langle\Delta\rho\rangle=\frac{1}{2}\int |\rho_{sDFT}(\br) - \rho_{KS-DFT}(\br)|d \br$.

\begin{table}[htp]
	\caption{\label{tab:density} MAD of the $\langle \Delta \rho \rangle^a$ computed with sDFT carried out with TF, revAPBEK, LMGPA and SMP NAKEs against KS-DFT results.}
	\begin{threeparttable}
		\begin{tabular}{cccccc}
			\hline
			\hline
			         Set           & Functional & Hydrogen & Dispersion & Mixed & Total \\ \hline
			\multirow{4}{*}{S22-5} &     TF     &   3.20   &    1.40    & 1.09  & 1.87  \\
			                       &  revAPBEK  &   3.07   &    1.67    & 0.96  & 1.89  \\
			                       &   LMGPA    &   3.07   &    1.45    & 0.90  & 1.79  \\
			                       &    SMP     &   2.95   &    1.14    & 0.89  & 1.64  \\ \hline
			 \multirow{4}{*}{S66}  &     TF     &   3.52   &    2.08    & 1.76  & 2.49  \\
			                       &  revAPBEK  &   3.29   &    2.17    & 1.60  & 2.39  \\
			                       &   LMGPA    &   3.30   &    1.82    & 1.47  & 2.23  \\
			                       &    SMP     &   3.15   &    1.36    & 1.33  & 1.97 \\ \hline
		\end{tabular}
\begin{tablenotes}
	\footnotesize
	\item[a] The actual value of $\langle \Delta \rho \rangle$ are the values in the table multiplied by $10^{-2}$.
\end{tablenotes}         
\end{threeparttable}
\end{table}

The table shows that even though all of the adopted functionals appear to deliver electron densities that are very close to the KS-DFT electron density (all values in the table should be multiplied by $10^{-2}$)  SMP outperforms all other functionals. Once again, all functionals and SMP in particular, do an excellent job with dispersion-bound complexes. However, they deviate more from the KS-DFT benchmark for hydrogen-bonded complexes.

\begin{figure}[htp]
	\begin{tabular}{cc}
		S22-5 & S66 \\
	\includegraphics[width=0.5\linewidth]{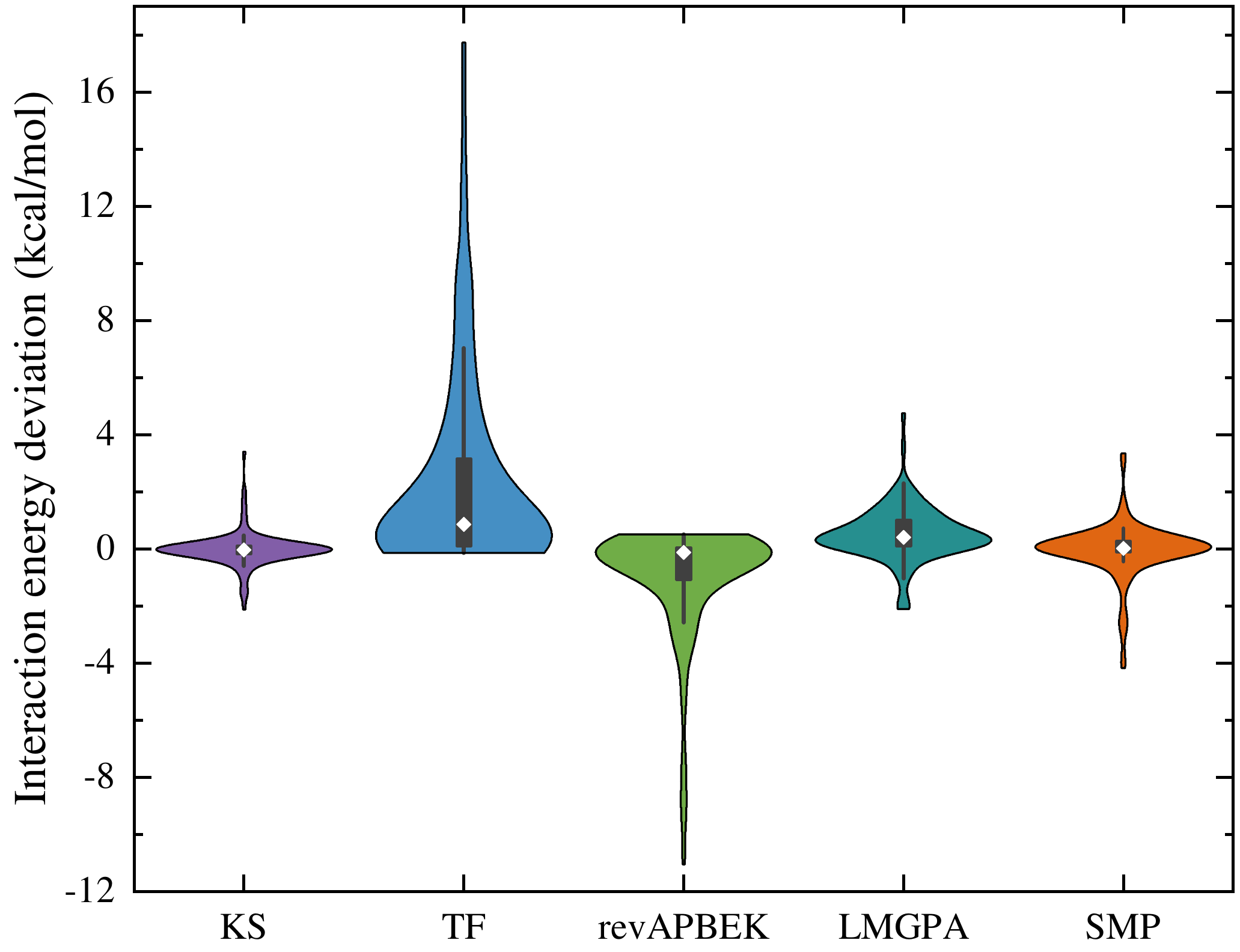} &
	\includegraphics[width=0.5\linewidth]{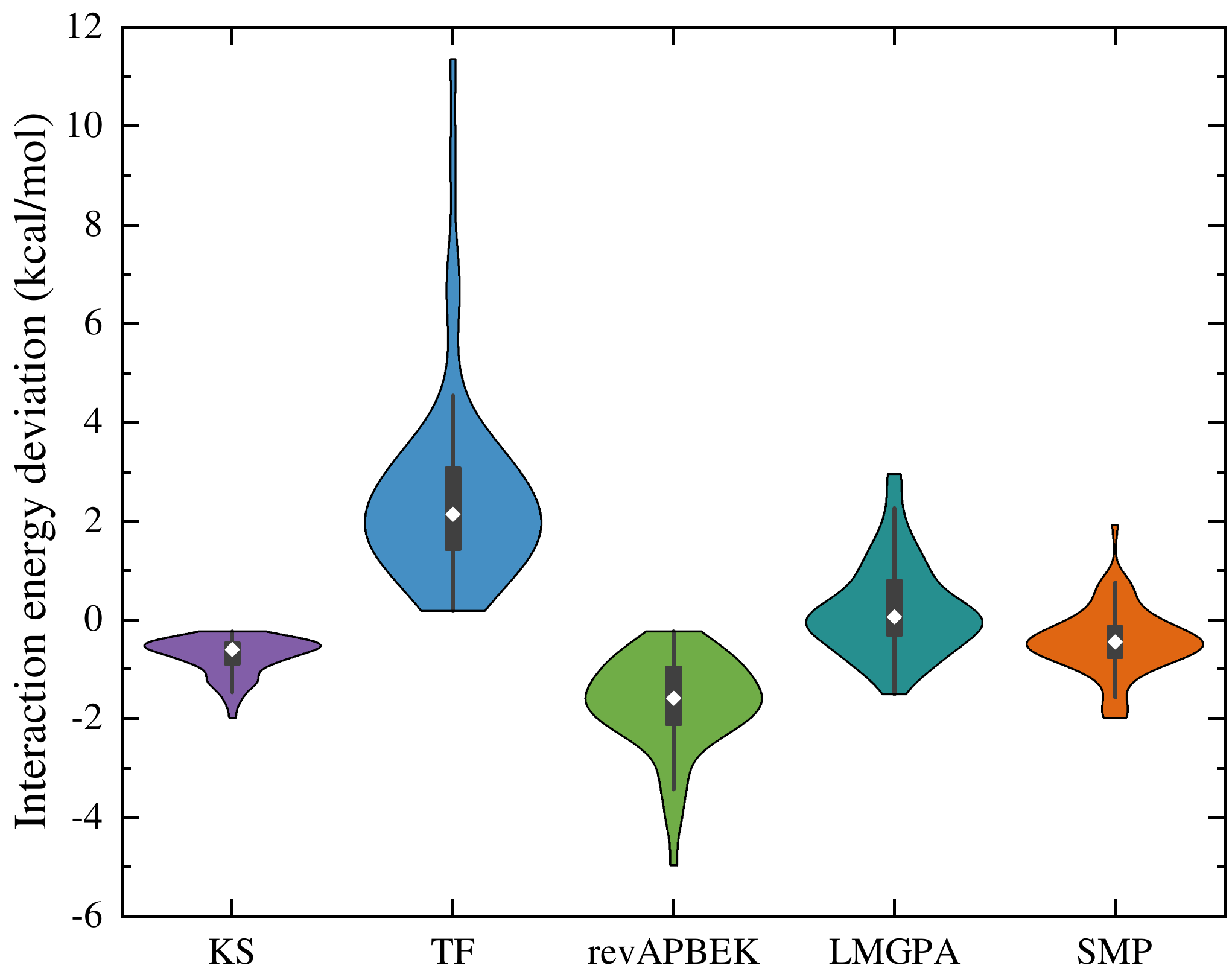}
	\end{tabular}
	\caption{\label{fig:ccsd} Violin plots overlaid on box plots of the interaction energy deviations of KS-DFT and sDFT with several NAKE functionals for the S22-5 and S66 test sets comparing against CCSD(T) interaction energies. Both KS-DFT and sDFT are carried out with PBE+D4 xc functional. The box (thick black bar) in the centre represents the interquartile range, and the thin black line indicate the range of data falling within 1.5 $ \times $ box-length beyond the upper and lower limits of the box. White diamond mark in the middle is the median value.}
\end{figure}

We conclude this section with an analysis of the performance of the NAKE functionals considered so far (including SMP) in production-like sDFT simulations by comparing the sDFT interaction energy results against benchmark CCSD(T) values. We choose the PBE+D4 xc functional as it was shown to perform particularly well for weak interactions while maintaining the cost of a semilocal functional \cite{Caldeweyher_2017,Caldeweyher_2019}. In Figure \ref{fig:ccsd} we show violin plots indicating that for both S22-5 and S66, SMP is the top performer, not only because the MAD are very close to zero (MAD$_{\rm S22-5}^{\rm SMP}=0.51$ kcal/mol, MAD$_{\rm S66}^{\rm SMP}=0.64$ kcal/mol.), but because the spread of the error is comparable to KS-DFT (for which the D4 corrections were parametrized) and evenly distributed (no bias). Conversely, similar to the results obtained when comparing against KS-DFT, TF and revAPBEK continue to show bias in opposite directions (TF overestimates and revAPBEK underestimates). The nonlocal LMGPA delivers interaction energy deviations that are slightly more deviated from the benchmark and spread out compared to SMP (MAD$_{\rm S22-5}^{\rm LMGPA}=0.77$  kcal/mol, MAD$_{\rm S66}^{\rm LMGPA}=0.69$ kcal/mol). 

\section{Conclusions}
This work proposes a new nonadditive kinetic energy functional to be used in subsystem DFT (or density embedding) calculations. The developed functional (termed SMP) is inspired by the role of the nonlocal part in nonlocal kinetic energy functionals, i.e., to effectively remove the \vw\ functional and only keep Thomas-Fermi in the high density regions and do the opposite in low density regions. SMP effectively interpolates from a GGA functional (we choose revAPBEK) to the Thomas-Fermi functional in high-density regions. Additionally, SMP is (sub)system-specific because it contains a parameter, $\rho^*$, determined automatically, and specific to the particular inter-subsystem density overlap occurring in the simulations. SMP is overall very accurate, improving on the two most accurate nonadditive kinetic energy functionals available to date for reproducing conventional DFT as well as benchmark CCSD(T) interaction energy values for the S22-5 and S66 test sets retaining the computational cost of a semilocal functional.

\section*{Acknowledgements}
This material is based upon work supported by the National Science Foundation under Grants No.\ CHE-1553993 and OAC-1931473. We thank the Office of Advanced Research Computing at Rutgers for providing access to the Amarel cluster. Xuecheng Shao acknowledges the Molecular Sciences Software Institute for support through a Software Investment Fellowship.

\bibliography{../prg_bibliography/prg}
\end{document}